\documentclass[desactivate]{aa}  

\begin{document}
\title{Forgotten treasures in the HST/FOC UV imaging polarimetric archives of active galactic nuclei}%
\subtitle{IV. 5 Orphaned AGNs}%
\titlerunning{Forgotten treasures in the HST/FOC UV imaging polarimetric archives -- IV. Orphaned AGNs}%
\author{T. Barnouin\inst{1}\thanks{thibault.barnouin@astro.unistra.fr}
    \and
    F. Marin\inst{1}
    \and
    E. Lopez-Rodriguez\inst{2,3}
}
\institute{Universit\'e de Strasbourg, CNRS, Observatoire astronomique de Strasbourg, UMR 7550, F-67000 Strasbourg, France
    \and
    Department of Physics \& Astronomy, University of South Carolina, Columbia, SC 29208, USA
    \and
    Kavli Institute for Particle Astrophysics \& Cosmology (KIPAC), Stanford University, Stanford, CA 94305, USA
}
\date{Received June 24, 2025; accepted August 5, 2025}
\abstract
{The Faint Object Camera (FOC) on board the Hubble Space Telescope (HST) acquired high-resolution spatially resolved polarimetric images of nearby active galactic nuclei (AGNs) in the near-ultraviolet (near-UV) band. Eight of the 25 individual targets in the polarized archives had no published analysis until the beginning of this series of papers. We describe the last 5 targets here.}
{In this paper, we finalize the publication of near-UV imaging polarimetry of AGNs in the HST/FOC archives. We render available spatially resolved polarization maps of the [OIII] emission lines for \object{Mrk 3} and \object{Mrk 78}, as well as near-UV continuum polarization maps for \object{Mrk 3}, \object{NGC 3862}, \object{Cygnus A}, and \object{3C 109}.}
{We used the generalized reduction pipeline presented in the first paper in this series to homogeneously analyze the five remaining polarized observations of AGNs in the FOC archives.}
{The polarization pattern in \object{Mrk 3} and \object{Mrk 78}in the narrow-line regions is consistent with scattering from an obscured nucleus. For \object{NGC 3862}, we confirm marginal UV polarization parallel with the inner radio jet that is related to synchrotron emission. In \object{Cygnus A}, we report spatially resolved centro-symmetric polarization patterns in the two opposite outflows, which highlights the scattering origin of the polarized light. Finally, \object{3C 109} shows high nuclear polarization that is consistent with AGN-dominated emission and parallel with the radio axis, but differs from the polarization from dichroic absorption invoked by previous authors.}
{The imaging polarimetry we obtained for the narrow-line region and the extended scattering medium surrounding the obscured AGNs is aligned with the predictions of the unified AGN model and demonstrates the power of spatially resolved polarimetric observation to decipher the complex morphologies at work in AGNs.}
\keywords{Instrumentation: polarimeters -- Methods: observational -- polarization -- Astronomical data bases: miscellaneous -- Galaxies: active -- Galaxies: Seyfert}
\maketitle
\section{Introduction}
\label{sec:Introduction}%
Active galactic nuclei (AGNs) are the bright and energetic region at the center of certain massive galaxies. They are powered by accretion onto a supermassive black hole. As gas and dust fall in, matter heats and emits large amounts of radiation that often outshine the entire host galaxy in which they reside. The plasma in the accretion flow is commonly assumed to be optically thick and thermal \citep{Lynden1969}. Under this assumption, a significant portion of the bolometric luminosity is expected to emerge as optical and ultraviolet (UV) radiation, the latter of which dominates the former. The spectral feature associated with this thermal release is often called the big blue bump (BBB), but its nature is far more complicated than what is implied by this toy model \citep{Antonucci2012}. Because the source of the emission is concentrated in a tiny volume (with a radius $\lesssim0.1$ light days; see, e.g., \citealt{Edelson1996}), however, we are still far from being able to routinely spatially resolve these regions, and therefore, from understanding the mechanisms of accretion and consequently of ejection, and the associated general relativity effects.

To overcome this problem, more and more sensitive imaging and spectroscopic instruments were sent into space, where UV radiation is less blocked or highly attenuated by the Earth's atmosphere. The Hubble Space Telescope (HST) was a key piece for the advancement of science and the progress of our understanding of the central regions of galaxies, and particularly those that contain AGNs that emit, scatter, or absorb UV light. At launch, the HST embarked two polarization-sensitive instruments, one spectrometer (the Faint Object Spectrograph, or FOS; \citealt{Harms1991}) and one high-angular resolution imager (the Faint Object Camera, or FOC; \citealt{Laurance1990}). They were used for several years to try to decipher how black holes accrete matter, grow through cosmic time, and influence their host galaxies \citep{Bowen1994,Zheng1997,Rao1998}. In addition, because AGNs are ideal background light sources, the FOS and FOC were used to study the intergalactic medium (IGM), the circumgalactic medium (CGM), the interstellar medium (ISM), and galactic halos (see, e.g., \citealt{Clayton1995,Kerber2002}). The FOC, in particular, was of prime importance to survey the outflows in nearby and intermediate-redshift AGNs with a high spatial resolution to ascertain the shape of the continuum in the extreme-UV and to study radiation reprocessing near black holes and their accretion disk \citep{Antonucci1994,Capetti1995b,Kishimoto1999}.

Even though AGN observations fulfilled multiple scientific objectives, the FOS and FOC were both decommissioned and replaced by other instruments. Because it is complex to reduce an observation with polarization, a large fraction of the observation were never fully analyzed or published. This means that the archives of the FOS and FOC, the instruments that interest us here, still contain many potentially interesting data. We undertook a series of numerical archaeological research in the Mikulski Archive for Space Telescopes (MAST) and extracted particularly interesting unpublished data on several AGNs. In \citet{Barnouin2023}, we presented a new pipeline in \textsc{python} that is able to reduce all the FOC observations in imaging polarimetry in a homogeneous way. We benchmarked it against \object{NGC 1068} and started our polarimetric investigation using a forgotten observation of \object{IC 5063}, which showed how complex the interactions between the AGN outflows and the surrounding ISM are. In the second paper of our series \citep{Barnouin2024}, we explored the unpublished data of \object{Mrk 463E} and determined the location of its obscured nucleus and the inclination of its polar winds and their ejection history with great precision through polarization mapping. In addition, our maps revealed a streamer that connects \object{Mrk 463E} and \object{Mrk 463W}, with a tentative detection of a large kiloparsec-scale ordered magnetic field that connects the two galaxies. The third paper of our series focused on the time-dependent polarimetric variations in the jet of \object{M87} with a a five-year monitoring HST/FOC campaign that was never analyzed in polarization \citep{Marin2024}.

This paper describes the five remaining AGN targets with associated FOC polarization data that were never fully analyzed or published. This is the penultimate publication before a complete re-reduction of the FOC AGN catalog with a single analysis pipeline and a statistical study of the measurements obtained. In Sect.~\ref{sec:Reduction}, we briefly summarize the data reduction with our pipeline and redefine the most important quantities that we used throughout the publication. In Sect.~\ref{sec:Sample} we present the sample in great detail and explore their associated polarization maps in the near-UV that were not published before. We highlight their morphology and physics and discuss the scientific potential of these archival observations in Sect.~\ref{sec:Discussion} before we conclude in Sect.~\ref{sec:Conclusion}.

\section{Data reduction}
\label{sec:Reduction}%
The polarized FOC data we analyzed were retrieved from the MAST HST Legacy Archive\footnote{\url{https://archive.stsci.edu/missions-and-data/hst}}. We then used the generalized reduction pipeline presented by \citet{Barnouin2023}, the first paper of this series. We only describe the specific parameters we used for data reduction of our sample here and refer to the aforementioned paper for technical details on the pipeline. MAST-calibrated POL0, POL60, and POL120 images for each observations were cropped to the usable field of view (FoV), without warped or null edges. The background was estimated by fitting a Gaussian plus polynomial on the intensity histogram of each image using the Scott sampling rule \citep{Scott1979}. We defined the background level to be the Gaussian mode plus one standard deviation. This background value was used as statistical uncertainty baseline and was subtracted from the observed flux. A cross-correlation with oversampling allowed us to align the images to a precision of \hbox{$0.1$ pixel} within each polarizer filter and to compute the registration uncertainties. To maximize the signal-to-noise ratio (S/N), and unless specified otherwise, the images were spatially resampled to \hbox{0.10\arcsec$\times$0.10\arcsec}\ and smoothed using a Gaussian kernel with a full-width at half-maximum (FWHM) of 0.15\arcsec, i.e. \hbox{$1.5$ times} the size of a resampled pixel. For each pixel, the Stokes parameters I, Q, U and their associated variance-covariance matrix were computed using eqs.~9 and 10 from \cite{Barnouin2023}. The obtained Stokes parameters and variance-covariance matrix were rotated so that north was up ($\Psi = 0^\circ$), following the IAU convention: The value for the electric-vector position angle of polarization starts from north and increases through east. Finally, the observed polarization degree $P_\text{obs}$ and angle $\Psi$ were computed from the Stokes parameters, and the associated uncertainties were propagated. No deconvolution was applied during the process.

The reported polarization was debiased following the recommendations by \citet{Simmons1985}, \hbox{$P = \sqrt{{P_\text{obs}}^2-{\sigma_P^\text{stat}}^2}$}. We add that we updated the pipeline\footnote{\url{https://git.unistra.fr/t.barnouin/FOC_Reduction.git}} since its publication such that debiased polarization is computed using the propagated statistical uncertainties instead of full uncertainties. Unless specified, polarization vectors are displayed for an S/N on the intensity greater than ten ($\left[S/N\right]_I \geq 10$) and with an S/N on the polarization degree greater than three ($\left[S/N\right]_P \geq 3$). For each polarization map, we overlaid the polarization vectors for which the length was proportional to the debiased polarization degree, oriented according to the polarization angle and IAU convention. Integrated values for each Stokes flux (and their associated uncertainties) were computed by summing all spatial bins (uncertainties were summed quadratically) falling in the region of interest. We then computed the integrated polarization components and report the debiased quantities.

\section{The sample}
\label{sec:Sample}%
\begin{table*}[!ht]%
    \caption{AGN sample.}%
    \centering%
    \begin{tabular}{c c c c c c}
        Name & Coordinates (J2000) & Heliocentric $z$ & Hubble Distance & Proposal ID (Filter) & Exposure time (s) \\%
        \hline\hline%
        \object{Mrk 3} & 93.901519\degr, 71.037525\degr\ & 0.013509 & 59.59~Mpc &%
        \begin{tabular}{c}%
             5918 (F502M)\\%
             6702 (F275W)$^1$\\%
             6702 (F342W)$^1$%
        \end{tabular}&%
        \begin{tabular}{c}%
             5262\\%
             5187\\%
             1697%
        \end{tabular}\\%
        \hline%
        \object{NGC 3862} & 176.270871\degr, 19.606317\degr\ & 0.021595 & 100.33~Mpc & 4833 (F342W) & 896\\%
        \hline%
        \object{Mrk 78} & 115.673881\degr, 65.177072\degr\ & 0.037983 & 168.74~Mpc & 5918 (F502M) & 9122\\%
        \hline%
        \object{Cygnus A} & 299.868153\degr, 40.733916\degr\ & 0.056200 & 245.32~Mpc &%
        \begin{tabular}{c}%
            3790 (F320W)$^2$\\%
            6510 (F275W)\\%
            6510 (F342W)%
        \end{tabular}&%
        \begin{tabular}{c}%
             602\\%
             14148\\%
             48789%
        \end{tabular}\\%
        \hline%
        \object{3C 109} & 63.418208\degr, 11.203833\degr\ & 0.306000 & 1351.57~Mpc & 6927 (F342W) & 603\\%
    \end{tabular}%
    \tablefoot{The five AGNs we studied, ordered by increasing distance. The coordinates are given in the J2000 equatorial system, the redshifts are heliocentric (i.e., the Earth's rotational and orbital motions were removed from the measured velocity), and the Hubble distances use the standard $\Lambda$CDM cosmology \citep[H$_0$ = 67.8 km/sec/Mpc, $\Omega_\text{matter}$ = 0.308, $\Omega_\text{vacuum}$ = 0.692 ;][]{Planck2015}. $1$: Polarization maps published by \cite{Kishimoto2002}, but reprocessed with our pipeline. $2$: Integrated polarization value published by \cite{Antonucci1994,Hurt1999}.}\label{tab:sample}\vspace{-6pt}%
\end{table*}%

The MAST HST/FOC polarized archives contain observations of 25 different AGNs, 8 of which were unpublished or had a partially analyzed polarization until this series of papers. The sample of observations analyzed in the following section consists of the 5 AGNs that remain partially or completely unpublished to this day. Table~\ref{tab:sample} details the targets and their associated observations.

HST program GO 5918 (PI Axon) requested the observation of the narrow-line region (NLR) of three famous AGNs: \object{IC 5063}, \object{Mrk 3}, and \object{Mrk 78}. In this section, we reduce and analyze the archival data for \object{Mrk 3} and \object{Mrk 78}, as \object{IC 5063} was analyzed in the first paper of this series \citep{Barnouin2023}. For comparison and completeness purposes, we reduced and analyzed observation ID 5918 of \object{Mrk 3} alongside the other two polarized observation of the same object obtained during Proposal ID 6702 (PI Kay) through filters F275W ($\lambda \sim$ 2800 \AA) and F342W ($\lambda \sim$ 3400 \AA) and published in \cite{Kishimoto2002}.

Three polarized observations exist for \object{Cygnus A} (3C~405). Proposal ID 3790 (PI Antonucci, through filter F320W ; $\lambda \sim$ 3200 \AA) occurred before the Corrective Optics Space Telescope Axial Replacement (COSTAR) installation in December 1993. This observation was partially published, in the sense that only the integrated polarization values are available in \cite{Antonucci1994} and \cite{Hurt1999}. Polarized data obtained post-COSTAR installation for \object{Cygnus A} during proposal ID 6510 (PI Antonucci, through filters F275W and F342W) were never published.

Finally, \object{NGC 3862} (proposal ID 4833, PI Crane) and \object{3C 109} (proposal ID 6927, PI Antonucci) polarization was observed to study their optical jet and off-nuclear scattering medium, respectively, but were never analyzed. In the following subsections, we present the polarization maps obtained for each AGN and their respective analysis, ordered by increasing distance from Earth.

\subsection{\object{Mrk 3}}
\label{subsec:Mrk3}%
\object{Mrk 3} (UGC~3426) is classified as a Seyfert 2 galaxy. Its optical spectra only show narrow (FWHM < 1000 km s$^{-1}$) emission lines. Studies of its polarized-light spectrum revealed broad emission lines from a Seyfert 1-like nucleus \citep{Schmidt1985,Miller1990,Tran1995}, which confirmed the basic predictions from the Unified Model \citep{Antonucci1993}. \object{Mrk 3} is a nearby AGN, with a redshift of \hbox{$z = 0.013509$} \citep{Falco1999}, residing at a distance of \hbox{$\sim$ 59 Mpc} (in standard $\Lambda$CDM cosmology). At this distance, 1\arcsec\ corresponds to a transverse size of \hbox{$\sim$ 290 pc}. Its closeness and brightness makes it a prime target for polarimetric studies at all wavebands.

\begin{figure*}[!ht]%
    \centering%
    \includegraphics[height=6cm]{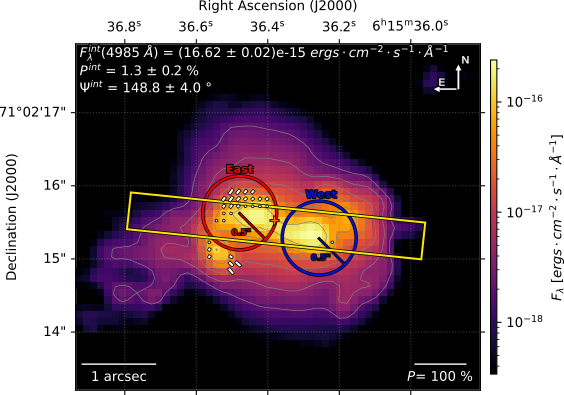}\hspace{4mm}%
    \includegraphics[height=6cm]{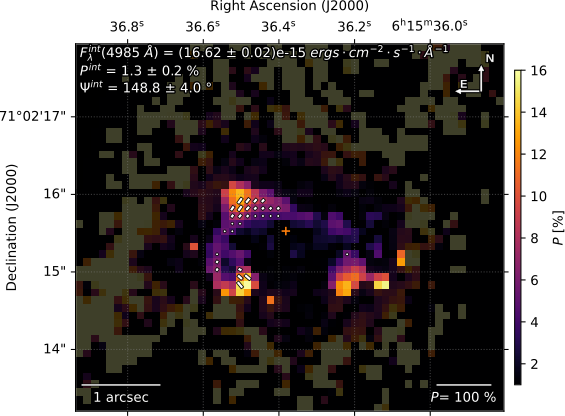}%
    \caption{HST/FOC polarization maps of \object{Mrk 3} through filter F502M. \textit{Left:} Intensity map overlaid with the integration regions used in Table~\ref{tab:Mrk3}. \textit{Right:} Debiased polarization degree. Darker pixels have $\left[S/N\right]_P < 1$. Both maps are at spatial binning of 0.10\arcsec, smoothed using a 0.15\arcsec\ FWHM Gaussian kernel. 30 pixels have $\left[S/N\right]_P \geq 3$. The orange cross displays the estimated location for the obscured nucleus \citep{Kishimoto2002}.}\label{fig:Mrk3_pol}\vspace{-6pt}%
\end{figure*}

\object{Mrk 3} is also famous for showing one of the clearest examples of a close association between radio and line emission. On one hand, radio observations show a linear structure that extends over \hbox{$\sim$ 2\arcsec} and is dominated by two symmetric and highly collimated jets at position angle \hbox{PA $\sim$ 84\degr} \citep{Kukula1993}. On the other hand, HST emission-line images, including [OIII] imaging presented in Fig. \ref{fig:Mrk3_pol}, reveal an NLR with an S-shaped morphology that is composed of a series of knots and filaments. It is basically co-spatial with the radio jets \citep[see also][]{Capetti1995}. This S-shape is also observed in radio \citep{Kukula1993} and might be caused by a change in the ejection axis of the jets with time or by the interaction of the jets with a rotating ISM.

Previous narrow-band [OIII] imaging of the NLR of Seyfert 1 and 2 galaxies, including \object{Mrk 3}, was obtained using the HST/FOC. The results were presented by \cite{SK96}: The authors estimated H$_\beta$ to cause 7.5\% of the total emission in the F502M filter, the remainder of the emission being dominated by the forbidden emission lines. The same [OIII] imaging shows a biconical emission-line region along \hbox{PA $\sim 70$\degr} with a full opening angle of the western cone of 40\degr\ for an extension of 1.4\arcsec\ (370pc) and a full opening angle of the eastern cone of 50\degr\ with and extension of 0.7\arcsec\ (185pc). More detailed studies of the kinematics of the NLR and extended NLR (ENLR) set the biconical NLR to \hbox{PA $\sim$ 89\degr}, confined to a narrow ridge with the line-emitting gas compressed by the shocks created by the passage of the radio jet \citep{Capetti1995}, while the biconical ENLR line emission is produced by unperturbed gas along \hbox{PA $\sim$ 112\degr} \citep{Ruiz2001,Crenshaw2010}.

Spectropolarimetry revealed the presence of broad H$_\alpha$ and H$_\beta$ emission lines in the polarized flux of \object{Mrk 3}, as described above, similar to the well-known example of a hidden Seyfert 1 in \object{NGC 1068} \citep{Schmidt1985,Miller1990}. The differences in the continuum polarization degree and in the rotation of the polarization angle suggest, however, that the scattering geometry of \object{Mrk 3} might be more complex than that of \object{NGC 1068}. Refining spectropolarimetry of \object{Mrk 3} over time allowed \cite{Tran1995} to uncover a continuum polarization of \hbox{$P \sim 7$\%} at \hbox{$\Psi \sim 167$\degr}, with an essentially wavelength-independent polarization from 3500\AA\ to 7000\AA\ and a rise to the blue of the polarized flux. This observation strengthened the electron-scattering hypothesis.

\begin{table*}[!ht]%
    \caption{Observed intensities and polarization from \object{Mrk 3} at various apertures drawn in Fig. \ref{fig:Mrk3_pol}.}\vspace{-6pt}%
    \centering\small%
    \resizebox{\linewidth}{!}{%
        \begin{tabular}{l|c c c|c c c|c c c}
             & \multicolumn{3}{c}{F275W} & \multicolumn{3}{c}{F342W} & \multicolumn{3}{c}{F502M} \\
            Aperture & Intensity & $P$ & $\Psi$ & Intensity & $P$ & $\Psi$ & Intensity & $P$ & $\Psi$ \\
            \hline
            FoV (\hbox{3\arcsec$\times$4\arcsec}) & 42.22 $\pm$ 0.13 & 14.2 $\pm$ 0.5 & 158.6 $\pm$ 1.0 & 34.50 $\pm$ 0.06 & 6.9 $\pm$ 0.3 & 164.7 $\pm$ 1.0 & 166.2 $\pm$ 0.2 & 1.3 $\pm$ 0.2 & 148.8 $\pm$ 4.0 \\
            \hbox{0.5\arcsec$\times$4\arcsec}\ slit & 25.67 $\pm$ 0.06 & 15.6 $\pm$ 0.4 & 156.6 $\pm$ 0.7 & 21.23 $\pm$ 0.04 & 7.7 $\pm$ 0.4 & 162.1 $\pm$ 1.1 & 103.1 $\pm$ 0.1 & 1.1 $\pm$ 0.2 & 136.3 $\pm$ 4.5 \\
            0.5\arcsec\ West & 14.76 $\pm$ 0.04 & 18.8 $\pm$ 0.5 & 158.2 $\pm$ 0.8 & 11.88 $\pm$ 0.03 & 10.1 $\pm$ 0.4 & 160.5 $\pm$ 1.2 & 57.53 $\pm$ 0.07 & 0.7 $\pm$ 0.2 & 151.1 $\pm$ 7.8 \\
            0.5\arcsec\ East & 11.40 $\pm$ 0.04 & 16.4 $\pm$ 0.5 & 144.0 $\pm$ 1.0 & 10.31 $\pm$ 0.03 & 7.3 $\pm$ 0.4 & 147.2 $\pm$ 1.8 & 51.21 $\pm$ 0.07 & 3.1 $\pm$ 0.2 & 131.6 $\pm$ 2.1 \\
        \end{tabular}%
    }\vspace{-6pt}%
    \tablefoot{Intensities are in units of 10$^{-16}$ erg~cm$^{-2}$~s$^{-1}$~\AA$^{-1}$, $P$ are in percent, and $\Psi$ are in degrees.}\label{tab:Mrk3}\vspace{-6pt}%
\end{table*}%
Following previous studies, we present in Fig.~\ref{fig:Mrk3_pol} a previously unpublished HST/FOC polarimetric image of \object{Mrk 3} (proposal ID 5918), obtained on February 27, 1998, after the installation of COSTAR. The observation used the F502M filter (530\AA\ bandwidth centered at 4940\AA), encompassing the essentially unpolarized \hbox{[O~III]$\lambda\lambda$4959,5007} doublet at the redshift of \object{Mrk 3}. A single exposure of 5\,261 seconds was taken for each polarizer.

The map in Fig.~\ref{fig:Mrk3_pol} shows the total flux and polarization degree, resampled to \hbox{0.10\arcsec$\times$0.10\arcsec} and smoothed with a Gaussian kernel of 0.15\arcsec\ FWHM. The polarization vectors are displayed where $\left[S/N\right]_P \geq 3$. We measure in the 3\arcsec$\times$4\arcsec\ FoV an integrated flux of \hbox{$F_\lambda(4940$\AA$) = (16.62 \pm 0.02) \times 10^{-15}$} erg\,cm$^{-2}$\,s$^{-1}$\,\AA$^{-1}$, with a polarization degree \hbox{$P = 1.3 \pm 0.2$\%} at \hbox{$\Psi = 148.8 \pm 4.0$\degr} (see Table~\ref{tab:Mrk3}). This is consistent with interstellar polarization (\hbox{$P_\text{ISP}\sim 1.2$\%} at \hbox{$\Psi_\text{ISP}\sim 132$\degr}, \citealt{Schmidt1985}). The small offset in angle suggests a contribution from scattered continuum emission.

The maps clearly reveal the S-shaped NLR, whose inner structure aligns with the radio axis at \hbox{PA = 84\degr}. While this central region is dominated by interstellar polarization, we detect a significant increase (up to 14\%) toward the outer edges of the NLR coma-like morphology.

To compare the polarization we obtained in this filter with previous HST/FOC observations, we integrated the polarized flux on synthetic apertures. We used a slit of width 0.5\arcsec\ and length 4\arcsec\ at position angle 84\degr\ (along the radio jet axis), and two circular apertures of radius 0.5\arcsec\ centered on the East and West NLRs, respectively. These apertures are drawn in the left panel of Fig. \ref{fig:Mrk3_pol}. The eastern part of the NLR displays the highest polarization degree (\hbox{$P = 3.2$\%} at \hbox{$\Psi = 132$\degr}), while the western NLR is almost undetected in polarization. This asymmetry might indicate that the eastern side is more strongly affected by scattering, possibly due to a more favorable geometry or a clearer line of sight. Alternatively, the western region might be subject to enhanced extinction or dilution by unpolarized emission, which would reduce the observed polarization signal. These spatial differences in the polarization structure are consistent with previous observations of Seyfert~2 galaxies (see, e.g., \object{NGC 1068}, \citealt{Capetti1995b}).

We also publish integrated polarization values for observation ID 6702, obtained on December 10, 1998, through two filters (F275W and F342W). We recall that the F275W filter is affected by Mg~II $\lambda$2800 emission and the F342W filter by [OII] $\lambda$3727, [NeV] $\lambda$3426 and [NeV] $\lambda$3346 emission lines. We integrated the polarization components on the same synthetic apertures as for the observation through the filter F502M, and we report them in Table \ref{tab:Mrk3} for comparison. Polarization maps were already published by \citet{Kishimoto2002}, but without quantitative values, other than the one of a resolved cloud near the nucleus location. In Figs. \ref{fig:MRK3_6702_F275W} and \ref{fig:MRK3_6702_F342W}, we show total intensity and debiased polarization degree maps of this observation, resampled to pixels of size \hbox{0.05\arcsec$\times$0.05\arcsec} and smoothed with a Gaussian with an FWHM 0.075\arcsec. They are substantially the same as those presented by \citet{Kishimoto2002}, and we therefore do not need to describe them in more detail. We discuss the measured polarization from the three maps, however, which was not done before. For the same aperture geometry, the polarization degree rises (from one percent to 15--20 \%) as the observed wavelength shifts to the UV. This reflects the diminishing impact of dilution by host starlight. Interestingly, the polarization degree of the western part, which was almost undetected in the F502M filter, becomes higher than that of the eastern part in the F342W and F275W filters. This is consistent with the hypothesis that the western and eastern NLRs are affected by different reddening.

\subsection{\object{NGC 3862}}
\label{subsec:NGC3862}%
\object{NGC 3862}, also known as \object{3C 264}, is an elliptical galaxy located in the galaxy cluster Abell~1367, residing at approximately 300 million light years away in the Leo constellation \citep{Smith2000}. At this distance, \hbox{1\arcsec\ $\sim$ 470~pc}. One of its most notable features is a prominent extragalactic jet associated with its central supermassive black hole, whose luminosity decreases as the distance from the central quasar host increases. This makes \object{NGC 3862} a Fanaroff-Riley type I (FRI) radio galaxy \citep{Fanaroff1974}. The jet was observed at multiple frequencies and scales and has an unresolved core and a smooth one-sided jet with evident variations in its morphological properties with distance that ends in a blob of emission at 28\arcsec\ from the core in the northeast direction \citep{Bridle1981,Gavazzi1981,Baum1988,Lara1997,Lara1999,Lara2004}.
\begin{figure*}[!ht]%
    \centering%
    \includegraphics[width=0.42\textwidth]{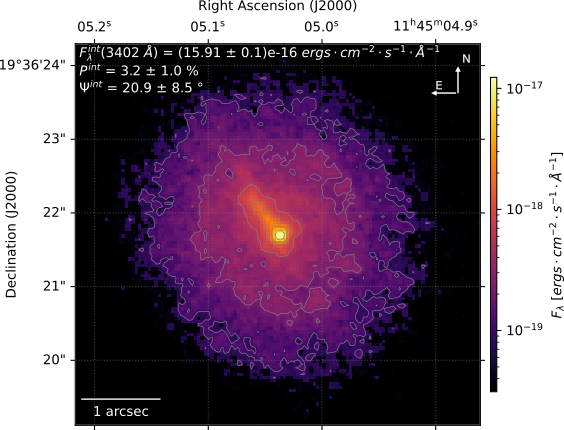}\hspace{4mm}%
    \includegraphics[width=0.42\textwidth]{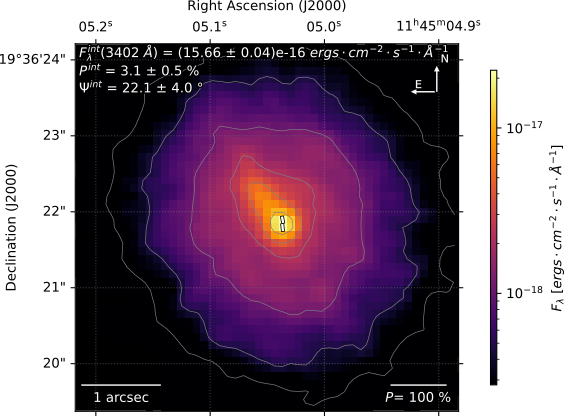}%
    \caption{HST/FOC total intensity map of \object{NGC 3862}, through filter F342W, cropped to the central 5\arcsec$\times$5\arcsec. \textit{Left:} Intensity map at the  Nyquist sampling resolution without smoothing. No polarization is detected among individual pixels. \textit{Right:} Intensity map at a spatial binning of 0.10\arcsec, smoothed using a 0.15\arcsec\ FWHM Gaussian kernel. 2 pixels have $\left[S/N\right]_P \geq 3$.}\label{fig:NGC3862_FOC}\vspace{-6pt}%
\end{figure*}%
The jet extends to about 0.6\arcsec\ in the optical band, so that the HST/FOC might resolve it in principle.. The source was observed by \citet{Crane1993,Crane1993b}, who indeed revealed the synchrotron emitting jet. The observation was made in the full \hbox{512$\times$512} pixel detector mode, using the F/96 camera with normal sized (unzoomed) pixels, providing a FoV of \hbox{11\arcsec$\times$11\arcsec} with a pixel size of 0.0224\arcsec. The source was observed by \citet{Crane1993b} on May 4, 1993 (before the COSTAR installation), for 896.75 seconds through each polarizer filter, once each. Interestingly, \citet{Crane1993b} only published the total intensity images of \object{NGC 3862}, reproduced in our Fig.~\ref{fig:NGC3862_FOC} (left). In our case, we used the latest calibration files, astrometry corrections, and reduction pipeline, which resulted in a sharper image and slightly different flux contours. The optical jet is clearly visible in the center of the image, where it extends to 0.61\arcsec\ in the northeast direction at a position angle of 38\degr. At a native pixel resolution, the map shows no clear evidence of polarization (less than a 2$\sigma_P$ detection of polarization integrated over the whole image).

We therefore decided to bin our map until $\geq 3\sigma_P$ polarization vectors were detected. This resulted in a lower-resolution map with a spatial binning of 0.10\arcsec, smoothed using a 0.15\arcsec\ FWHM Gaussian kernel. Even then, only two pixels have $\left[S/N\right]_P \geq 3$ (see Fig.~\ref{fig:NGC3862_FOC}, right). The integrated polarization over the whole FoV is \hbox{$P = 2.7 \pm 0.6$\%}, which allows us to claim a real detection (and measurement) of the 3400\AA\ polarization in \object{NGC 3862}. Integrating the polarization signal at various, decreasing circular aperture radii (from the full FoV to 0.5\arcsec), centered on the brightest intensity spot, does not change the orientation of the measured polarization angle, but increases the polarization degree from \hbox{$P = 2.7 \pm 0.6$\%} to \hbox{$P = 4.8 \pm 0.8$\%}, as less and less host starlight is diluting the polarized signal.

The associated aperture-independent polarization position angle is \hbox{$\Psi = 17.2 \pm 6.2$\degr}, which does not align with the jet position angle observed in the optical because, as already mentioned, the jet position angle evolves with distance from the unresolved core. European VLBI Network (EVN) 5~GHz maps have revealed that the base of the jet, at a few tenths of milliarcseconds, propagates along a position angle of 27\degr before it rotates at several hundred milliarcseconds \citep{Lara1997}. The near-UV polarization measured from our HST/FOC map is better aligned with the innermost jet position angle, implying that the polarization we measure comes from the AGN core rather than from scattering or dichroic absorption in the host galaxy itself. Because no broad emission lines are detected in the optical spectrum of this source even after host starlight removal \citep{Buttiglione2009}, the observed polarization is much higher than 1\% and is parallel to the radio axis, which strongly favors a synchrotron origin rather than scattering inside the disk or BLR.

If the polarized emission is indeed mainly related to synchrotron emission, then the polarization angle traces the orientation of the magnetic fields. In this case, the magnetic fields are perpendicular to the jet. Coupled to $P$ 3-5\%, we can explain the observed polarization of synchrotron emission by invoking relativistic reconfinement shocks with ordered magnetic fields, as shown by \citet{Nalewajko2012}. In these shocks, toroidal fields create polarization of $\sim$ 3\% with a parallel polarization angle, when the viewing angle is about 30$^\circ$ from the symmetry axis of the AGN. The deviation of about 10$^\circ$ with respect to a perfect alignment might be caused by scattering of optical radiation at the base of the NLR, which would produce a polarization angle perpendicular to the radio axis and would ultimately modify the total (synchrotron plus scattering) polarization angle.

\subsection{\object{Mrk 78}}
\label{subsec:Mrk78}%
Mrk 78 is a Seyfert 2 galaxy at redshift of \hbox{$z = 0.037983$} \citep{Koss2022}, or approximately \hbox{$\sim 168$ Mpc} in standard $\Lambda$CDM cosmology. At this distance, 1\arcsec\ corresponds to a transverse size of \hbox{$\sim 785$ pc}. \object{Mrk 78} has been extensively studied for its strong correlation between its observed radio components, its thick dust lane, and its extended NLR. Observations in radio show a triple radio source that can be attributed to a nucleus and the eastern and western components, aligned on radio \hbox{PA $\sim 90$\degr} \citep{Ulvestad1981,Pedlar1989}. High-resolution imaging of the [OIII] line-emission of \object{Mrk 78} with HST/FOC revealed a biconical NLR at position angle 67\degr, slightly misaligned with the radio extension, with a full opening angle of 40\degr\ and an extension of 4\arcsec\ ($\sim3000$~pc, \citealt{Capetti1994,Capetti1996,SK96}). Further observations in the radio and UV allowed \cite{Whittle2004} to study the jet-gas interactions. They found that because the radio source material and line-emitting gas do not easily interpenetrate, radio jet flows can be disrupted and deflected by the gas. They can also distort or ablate the line-emitting gas as well as coherently accelerating it. \cite{Whittle2004} noted that \object{Mrk 78} showed a sequence of initial, intermediate, and late stages of an evolutionary sequence for the jet-gas interaction. This post-merger object also displays double-peaked [OIII] emission lines \citep{Whittle1988,Nelson1995}, which fueled the discussion whether a pair of supermassive black holes might lie at its core, but \citep{Fischer2011} showed that this feature can also be explained by an asymmetric distribution of the out-flowing gas in the NLR.

In an asymmetric system like this, we expect high detectable polarization. but previous polarimetric campaigns of \object{Mrk 78} are scarce. Only two results were published. The first result reported \hbox{$P \sim 0.1$\%} at \hbox{$\Psi \sim 14$\degr}\ in the optical band, but the measurements are too noisy to be conclusive \citep{Miller1990,Kay1994}. The second campaign reported \hbox{$P = 0.25 \pm 0.09$\%} at \hbox{$\Psi = 168 \pm 10$\degr}\ in the [OIII] narrow-line emission \citep{Goodrich1992b}, which holds no clues about the behavior of the continuum polarization.
\begin{figure*}[!ht]%
    \centering%
    \includegraphics[height=4.5cm]{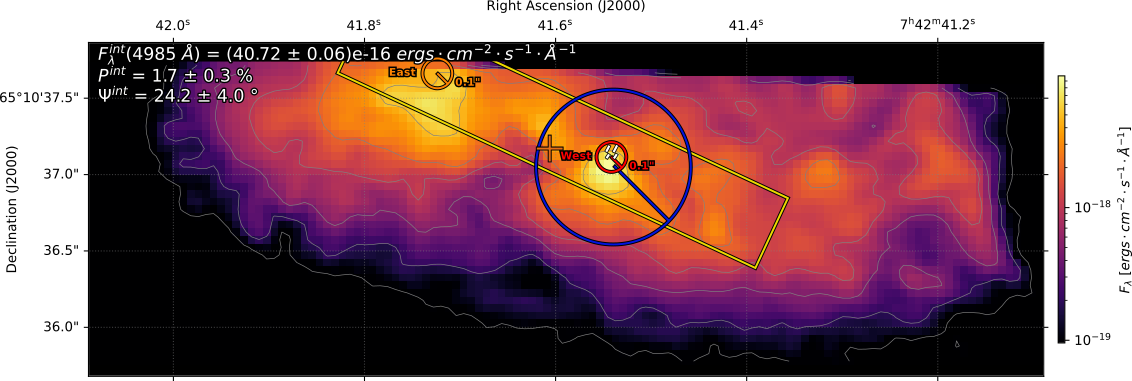}\\\vspace{2mm}%
    \includegraphics[height=4.5cm]{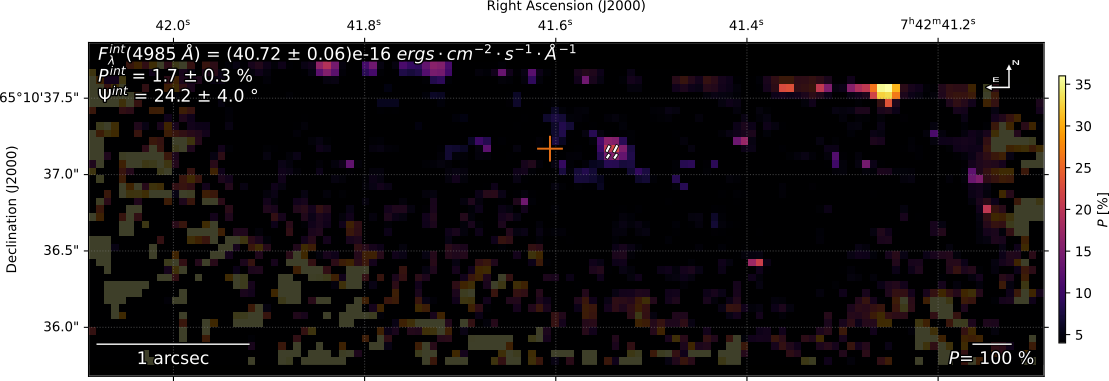}%
    \caption{HST/FOC polarization maps of \object{Mrk 78} through filter F502M. \textit{Top:} Intensity map overlaid with the integration regions used in Table~\ref{tab:MRK78}. \textit{Bottom:} Debiased polarization degree. Darker pixels have $\left[S/N\right]_P < 1$. Both maps are at a spatial binning of 0.05\arcsec, smoothed using a 0.075\arcsec\ FWHM Gaussian kernel. 4 pixels have $\left[S/N\right]_P \geq 3$. The orange cross displays the estimated location for the obscured nucleus \citep{Whittle2004}.}\label{fig:Mrk78_pol}\vspace{-6pt}%
\end{figure*}%
To explore this is greater detail, we present the observation ID 5918 of \object{Mrk 78}, obtained on September 12, 1997, through filter F502M. Unfortunately, \object{Mrk 78} is not centered in the FoV (see Fig.~\ref{fig:Mrk78_pol}, top), and we therefore cannot perform a proper analysis of the northeast part of the AGN that is outside or too close to the border of the FoV. We therefore focused on the western part of the NLR, which might display the initial and late stages of the jet-gas interactions \citep{Fischer2011}. We present in Fig.~\ref{fig:Mrk78_pol} the polarization maps resampled to spatial bins of size 0.05\arcsec, smoothed with a Gaussian kernel with an FWHM of 0.075\arcsec\, which gives us sufficient statistics to detect a polarized knot next to the western hotspot in the NLR. The obtained [OIII] image reproduces the prominent biconical NLR well along the position angle 67\degr\ observed in previous [OIII] observations \citep{Capetti1994,Capetti1996,SK96}. Fig.~\ref{fig:Mrk78_pol} (top) shows most of the NLR of \object{Mrk 78} and two bright spots separated by 1.25\arcsec\ ($\sim 980$~pc) bisected by the dust lane along a position angle 69\degr. The NLR moves beyond the FoV after the hotspot in the northeast direction, and it appears to turn after the southwest hotspot toward a \hbox{90--100\degr}\ position angle. From the polarization map analysis, we are able to report a polarization detection right north of the western hotspot (see Fig.~\ref{fig:Mrk78_pol}, top) with polarization in a single spatial bin as high as \hbox{$P = 22.20 \pm 4.86$\%} and integrated polarization as high as $11.8$\% in a simulated circular aperture of radius 0.1\arcsec\ (see Table~\ref{tab:MRK78}). These values are aligned with what we expect from a perpendicular scattering of the core photons onto the NLR (see, e.g., \citealt{Capetti1995b,Barnouin2023}).
\begin{table}[!ht]%
    \caption{Observed intensities and polarization from \object{Mrk 78} at various apertures drawn in Fig. \ref{fig:Mrk78_pol}.}\vspace{-6pt}%
    \centering%
    \resizebox{\linewidth}{!}{%
        \begin{tabular}{l|c c c}
            Aperture & Intensity & $P$ & $\Psi$ \\
            \hline
            FoV (\hbox{3\arcsec$\times$5\arcsec}) & 40.72 $\pm$ 0.06 & 1.7 $\pm$ 0.3 & 24.2 $\pm$ 4.0 \\
            \hbox{0.5\arcsec$\times$3\arcsec}\ slit & 14.97 $\pm$ 0.03 & 1.3 $\pm$ 0.4 & 171.7 $\pm$ 6.5 \\
            0.5\arcsec\ West & 8.28 $\pm$ 0.02 & 1.7 $\pm$ 0.5 & 144.3 $\pm$ 6.6 \\
            0.1\arcsec\ West & 6.41 $\pm$ 0.07 & 11.8 $\pm$ 1.5 & 149.8 $\pm$ 3.7 \\
            0.1\arcsec\ East & 4.78 $\pm$ 0.06 & 8.3 $\pm$ 2.0 & 49.4 $\pm$ 5.3 \\
        \end{tabular}%
    }%
    \tablefoot{Intensities are in units of 10$^{-16}$ erg~cm$^{-2}$~s$^{-1}$~\AA$^{-1}$, $P$ are in percent, and $\Psi$ are in degrees. If the eastern aperture is above the detection level, it should be taken with caution as it lies right next to the edge of the FoV and might be affected by border effects.}\label{tab:MRK78}%
    \vspace{-6pt}%
\end{table}%
We detected a polarized cloud right north of the southwest hotspot. An integration of the Stokes Q and U over 0.10\arcsec\ aperture radius where the polarized signal is above $3\sigma_P$ yields \hbox{$P = 11.8 \pm 1.5$\%} at \hbox{$\Psi = 149.8 \pm 3.7$\degr}. This high polarization can be explained by photon scattering in a zone of higher NLR density. The photon source direction may then be found at the normal to the electrical polarization vector. In this scattering scenario, the obscured nucleus is consistent with being the obscured source of radiation. We also obtained a marginal detection of polarization right north of the northeast hotspot. Integrating Stokes Q and U over a similar aperture yields \hbox{$P = 8.3 \pm 2.0$\%} at \hbox{$\Psi = 49.4 \pm 5.3$\degr}. In this region, an electrical polarization vector at an angle of $\sim 50$\degr\ is almost perpendicular to the direction toward the nucleus. These values must be taken with caution, however, as proximity to the border of the detector can bring some border effects into the polarization reduction.

To compare our data with the previously published polarization values mentioned before, we integrated the polarization components detected in our map using a simulated slit with a size of \hbox{0.5\arcsec$\times$3\arcsec} at position angle 65\degr\ (along the NLR position angle). We also integrated the polarization on the full FoV and used circular apertures with different radii centered on the western and eastern hotspots. The results are shown in Table~\ref{tab:MRK78}. The emission lines are essentially polarized by transmission through aligned dust grains \citep{Antonucci1985} that can amount to \hbox{$P \sim 1.7$\%} at $\Psi \sim 27$\degr. We dismiss interstellar polarization in our galaxy due to its very low contribution in the direction of \object{Mrk 78} \citep[BD+68 496, \hbox{$P_\text{ISP} = 0.224 \pm 0.037$\%} at \hbox{$\Psi_\text{ISP} = 9 \pm 5$\degr}\ ; ][]{Panopoulou2025}. The lower polarization degree and different angle obtained in the slit indicates a polarized NLR at a different polarization angle. This polarization angle, along with those detected in the east and west polarized sources, is coherent with scattering of nuclear light on the NLR.

\subsection{\object{Cygnus A}}
\label{subsec:CygnusA}%
Cygnus A is a close radio galaxy \citep[z = 0.0562,][]{Horton2020} at 1\arcsec\ $\sim 1$ kpc, and it is one of the most frequently studied FR~II radio galaxies \citep{Carilli1996} because of its proximity and extreme characteristics. Its central few kiloparsec contain a patchy dust lane, a dusty ionization cone, and radio jets with a heavily extinguished core \citep{Taylor2003,Carilli2019,Sorathia1996,Lopez-Rodriguez2018}. High angular resolution observations of these components were not able to distinguish a dominating emission process between thermal and synchrotron components \citep{Privon2012,Koljonen2015}.

This is where polarimetric analysis can help by providing additional observables and constraints to the AGN core and its surroundings. Cygnus A has been extensively studied in polarimetry in the past, and revealed, among other discoveries, that the dominant polarization mechanism for optical and UV radiation arises from scattering \citep{Tadhunter1990,Jackson1993,Antonucci1994,Ogle1997,Hurt1999,Tadhunter2000}. In the near-IR, strong polarization perpendicular to the jet still shows evidence for scattering \citep{Packham1998,Ramirez2014}, while in the mid-IR, a rotation of the polarization angle highlights a change in the dominant polarization mechanism from scattering to dichroic extinction in the far-IR \citep{Lopez-Rodriguez2014,Lopez-Rodriguez2018}. The relativistic beaming of synchrotron emission into the plane of the sky, combined with the lack of temporal variability in the optical polarization, rules out synchrotron emission as the origin of the continuum polarization \citep{Jackson1993,Hurt1999,Bemmel2003}.

\begin{figure*}[!ht]%
    \centering%
    \includegraphics[height=6cm]{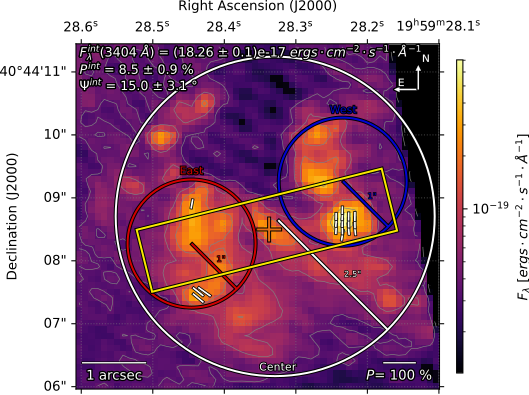}\hspace{4mm}%
    \includegraphics[height=6cm]{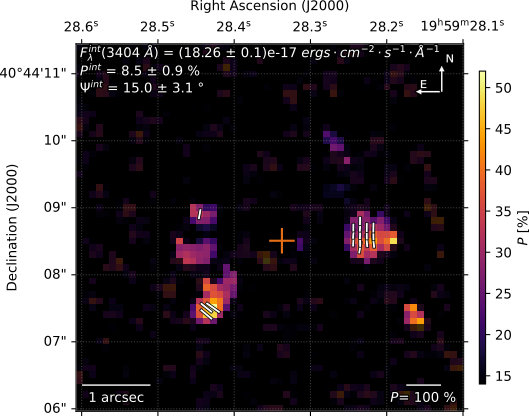}%
    \caption{HST/FOC polarization maps of \object{Cygnus A}, observation ID 6510 through filter F342W. \textit{Left:} Intensity map overlaid with the integration regions used in Table~\ref{tab:CygnusA}. \textit{Right:} Debiased polarization degree. Darker pixels have $\left[S/N\right]_P < 1$. Both maps have a spatial binning of 0.10\arcsec, smoothed using a 0.15\arcsec\ FWHM Gaussian kernel. 18 pixels have $\left[S/N\right]_P \geq 3$. The orange cross displays the estimated location for the obscured nucleus \citep{Jackson1998}.}\label{fig:CygnusA6510_pol}\vspace{-6pt}%
\end{figure*}

Our work allowed us to publish three previously unpublished polarization maps of \object{Cygnus A}. The first observation is ID 3790, obtained through filter F320W, acquired on October 28, 1992, that is, before the COSTAR installation. It is analyzed in Sect.~\ref{subsubsec:CygnusA3790}. The second map, ID 6510, was obtained through filter F342W on September 1, 1998, and it is explored in Sect.~\ref{subsubsec:CygnusA6510}. The third and final map, observation ID 6510 obtained through filter F275W on September 11, 1998, is discussed in Appendix.~\ref{app:CygnusA} because its S/N is very low. The first two observations benefited from enough integration time to allow us to perform a spatially resolved analysis. The integration regions displayed in Fig.~\ref{fig:CygnusA6510_pol} were used for the two observations and were compared to equivalent integration regions used by \cite{Hurt1999} in Table~\ref{tab:CygnusA}. The third observation was obtained with filter F275W ($\lambda2805$\AA) and had a total exposure time of 14\,148 seconds. \object{Cygnus A} is underexposed, unfortunately. No analysis could be performed on this dataset. The $\left[S/N\right]_I$ and $\left[S/N\right]_P$ maps are shown in Fig.~\ref{fig:CygnusA6510_lowflux} as a reference.
\begin{table*}[!ht]
    \caption{Observed intensities and polarization from \object{Cygnus A} at various apertures drawn in Fig. \ref{fig:CygnusA6510_pol}.}\vspace{-6pt}%
    \centering\small%
    \resizebox{\linewidth}{!}{%
        \begin{tabular}{l | c c c | c c c | c c c}%
             & \multicolumn{3}{c}{\cite{Hurt1999}} & \multicolumn{3}{c}{F320W} & \multicolumn{3}{c}{F342W} \\
            Aperture & Intensity & $P$ & $\Psi$ & Intensity & $P$ & $\Psi$ & Intensity & $P$ & $\Psi$ \\
            \hline
            FoV (\hbox{5\arcsec$\times$5\arcsec}) & - & - & - & 17.71 $\pm$ 0.18 & 6.3 $\pm$ 1.6 & 4.9 $\pm$ 7.1 & 18.26 $\pm$ 0.10 & 8.5 $\pm$ 0.9 & 15.0 $\pm$ 3.1 \\
            \hbox{1\arcsec$\times$4\arcsec}\ slit & - & - & - & 2.89 $\pm$ 0.06 & 9.7 $\pm$ 2.9 & 0.2 $\pm$ 7.7 & 5.27 $\pm$ 0.04 & 16.7 $\pm$ 1.2 & 13.3 $\pm$ 2.1 \\
            2.5\arcsec\ Center$^1$ & 7.67 & 5.7 $\pm$ 1.5 & 22 $\pm$ 8 & 9.53 $\pm$ 0.11 & 6.9 $\pm$ 1.8 & 13.5 $\pm$ 7.1 & 14.50 $\pm$ 0.08 & 10.0 $\pm$ 0.9 & 17.0 $\pm$ 2.7 \\
            1\arcsec\ West$^2$ & 1.47 & 12.6 $\pm$ 2.7 & 13 $\pm$ 6 & 2.01 $\pm$ 0.05 & 14.8 $\pm$ 3.6 & 10.6 $\pm$ 6.2 & 3.73 $\pm$ 0.04 & 17.7 $\pm$ 1.5 & 10.9 $\pm$ 2.4 \\
            1\arcsec\ East$^3$ & 3.34 & 2.9 $\pm$ 1.8 & 38 $\pm$ 18 & 2.06 $\pm$ 0.05 & (5.3 $\pm$ 3.6) & (47.5 $\pm$ 18.9) & 3.58 $\pm$ 0.04 & 14.1 $\pm$ 1.6 & 30.0 $\pm$ 3.0 \\
        \end{tabular}%
    }%
    \tablefoot{Intensities are in units of 10$^{-17}$ erg~cm$^{-2}$~s$^{-1}$~\AA$^{-1}$, $P$ are in percent, and $\Psi$ are in degrees. The values between parentheses are below the 3$\sigma$ level. The values from \cite{Hurt1999} from synthetic apertures are $1$) \hbox{5.65\arcsec$\times$5.65\arcsec}, $2$) \hbox{1.47\arcsec$\times$2.22\arcsec}, and $3$) \hbox{2.35\arcsec$\times$3.32\arcsec}.}\label{tab:CygnusA}%
    \vspace{-6pt}%
\end{table*}%

\subsubsection{Pre-COSTAR observation at $\lambda3112$~\AA}
\label{subsubsec:CygnusA3790}
While we can perform synthetic aperture integration to study the map, it remains underexposed, and we were unable to detect polarization in spatial bins as large as \hbox{0.7\arcsec$\times$0.7\arcsec} (50$\times$50 native pixels binning). This explains why only integrated polarization values were reported previously and no map was published \citep{Antonucci1994,Hurt1999}. A previous analysis of this dataset by \cite{Hurt1999} found \hbox{$P = 5.7 \pm 1.5$\%} at \hbox{$\Psi = 22 \pm 8$\degr} in a synthetic aperture with a size of \hbox{5.65\arcsec$\times$5.65\arcsec}, which confirmed scattering of nuclear light from a hidden quasar suggested by \cite{Ogle1997}. By integrating over smaller apertures on the northwest and southeast nuclei, the authors estimated that the polarized flux comes from the northwest nucleus. We compared the integrated polarization components from \cite{Hurt1999} to our own in Table~\ref{tab:CygnusA}. The map we obtained benefited from the latest recalibration of the HST/FOC archives in 2006 \citep{FOCcalib2006}and from the complete reprocessing through the pipeline. We are able to confirm values presented in \cite{Hurt1999} within the uncertainties: integrating over a 2.5\arcsec\ aperture radius centered on the nucleus gives us \hbox{$P = 6.9 \pm 1.8$\%} at \hbox{$\Psi = 13.5 \pm 7.1$\degr}. Integrating over a synthetic 1\arcsec\ aperture radius centered on the brightest spot northwest of the nucleus yields \hbox{$P = 14.8 \pm 3.6$\%} at \hbox{$\Psi = 10.6 \pm 6.2$\degr}. The observation lacks statistics to further study the AGN at $\lambda3112$\AA\ because we lack a detection of the integrated polarization in the southeast hotspot. The polarization obtained in the northwest source is coherent with polarization obtained on the whole polarized source, however, which confirms a dominant polarization from the western region. A slit with a size of 1\arcsec$\times$4\arcsec\ at a position angle 104\degr\ is also simulated along the radio jet axis. Its integrated polarization values are similar to the total aperture values within the uncertainties, and it is also coherent with a dominant contribution from the northwest polarized source.

\subsubsection{Post-COSTAR observation at $\lambda3404$~\AA}
\label{subsubsec:CygnusA6510}
Observation ID 6510 through filter F342W delivered the first exploitable polarization map of \object{Cygnus A} in near-UV with the great resolution of the HST/FOC instrument and a good exposure. With a total exposure time of 48\,790 seconds, we were able to produce polarization maps with a spatial binning of 0.10\arcsec, smoothed with a Gaussian kernel with an FWHM of 0.15\arcsec. We report a polarization detection in 18 spatial bins. We present in Fig.~\ref{fig:CygnusA6510_pol} the obtained polarization maps, with polarization vectors overlaid on the intensity map (left) and the polarization degree (right). In the southeast source, we detect a debiased polarization degree as high as \hbox{$P = 51.23 \pm 16.36$\%} ($3.13\ \sigma_P$). The few displayed polarization vectors show the formation of a centro-symmetric pattern, as observed in the V band by \cite{Tadhunter1990}. We integrated the polarization components in the same regions as for the pre-COSTAR observation analyzed previously and delimited on the total intensity map in Fig.~\ref{fig:CygnusA6510_pol} (left). The integrated values were compared to those obtained through F320W in the pre-COSTAR observation by \cite{Hurt1999} and this work in Table~\ref{tab:CygnusA}. The obtained integrated measures show a higher polarization degree than in the F320W band, which is expected because the dilution by the host galaxy at bluer wavelengths is lower. The statistics allowed us to study the polarization in the southeast and northwest regions, with \hbox{$P = 14.1 \pm 1.6$\% at $\Psi = 30.0 \pm 3.0$\degr} and \hbox{$P = 17.7 \pm 1.5$\% at $\Psi = 10.9 \pm 2.4$\degr}, respectively, in a synthetic 1\arcsec\ radius aperture. The differing polarization angle can be explained by the differing viewing angle of the obscured nucleus from each polarized region. The better statistics obtained in this observation show that the previously synthesized 1\arcsec$\times$4\arcsec\ slit along the radio jet axis contains the northwest polarized hotspot. Its integrated polarization components are also consistent with dominant contribution from the northwest polarized source, \hbox{$P = 16.7 \pm 1.2$\% at $\Psi = 13.3 \pm 2.1$\degr}.

For a better comparison with the polarization maps obtained in the $V$, $B$, and $K$ band \citep[respectively]{Tadhunter1990,Ogle1997,Packham1998}, we publish in Fig.~\ref{fig:CygnusA6510_largebins} the same observation resampled to large bins of 0.28\arcsec$\times$0.28\arcsec\ (corresponding to a rebinning of 20$\times$20 native pixels). The obtained polarization vectors show a clear centro-symmetric pattern that is consistent with the estimated location for the obscured nucleus \citep{Jackson1998}. The polarization degree map concurs on the provenance of polarized flux from the two hotspots observed in Fig.~\ref{fig:CygnusA6510_pol}. Comparing the total intensity and polarization degree maps in Figs.~\ref{fig:CygnusA6510_pol} and \ref{fig:CygnusA6510_largebins} to H$_\alpha$ and [OI] observations in \cite{Jackson1998} Figs.~2 and 4, we also observe an emitting region with paraboloidal boundary. Interestingly, we detect no polarization above the noise level where the radio jets seem to travel through the southeast and northwest winds. We report an average half-opening angle of $\sim40$\degr\ and $\sim50$\degr\ for the southeast and northwest cones, respectively.

\subsection{\object{3C 109}}
\label{subsec:3C109}%
\begin{figure*}[!ht]%
    \centering%
    \includegraphics[height=6cm]{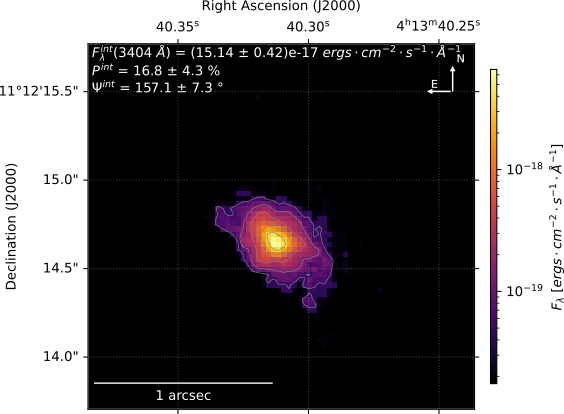}\hspace{4mm}%
    \includegraphics[height=6cm]{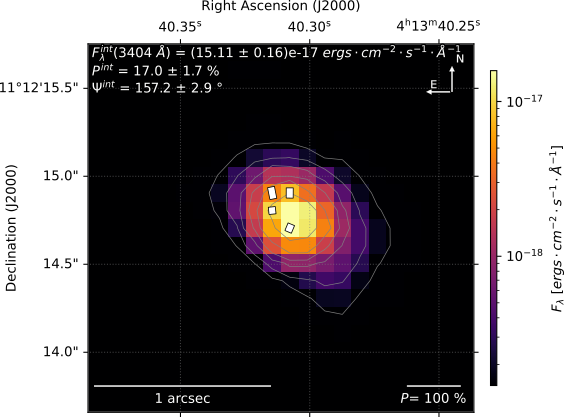}%
    \caption{HST/FOC total intensity maps of \object{3C 109}, observation ID 6927 through filter F342W, cropped to the central \hbox{2\arcsec$\times$2\arcsec}. \textit{Left:} Intensity map at the native spatial resolution without smoothing. No polarization is detected among individual pixels. \textit{Right:} Intensity map at a spatial binning of 0.10\arcsec, smoothed using a 0.15\arcsec\ FWHM Gaussian kernel. 4 pixels have $\left[S/N\right]_P \geq 3$. The maps are zoomed in, the color-bar flux values and scale of the pixels differ between the two figures.}\label{fig:3C109_FOC}\vspace{-6pt}%
\end{figure*}%
\object{3C 109} is the fifth and last AGN to be explored in this paper. It is a powerful broad-line radio galaxy \citep{Grandi1978} whose Eddington ratio was estimated to be super-Eddington \citep{Chalise2020}. This makes it an interesting target to understand accretion processes better. Observations in X-rays \citep{Allen1997} and IR \citep{Rudy1999} revealed that this accretion is unfortunately dimmed by intervening dust that might either be located around the accretion structure (the so-called circumnuclear torus in the classical interpretation of the Unified Model of AGN; see \citealt{Antonucci1993}) or inside an extended dust lane in the host galaxy. \cite{Goodrich1992} observe a large Balmer decrement of the broad lines, which confirms the presence of reddening by dust and would argue for the second option. This was confirmed by spectropolarimetry, where the continuum and broad lines of \object{3C 109} were found to be $\sim 8$\% polarized, probably by transmission through aligned dust grains within the host galaxy itself \citep{Goodrich1992}. Ejection and feedback also occur in \object{3C 109} in the form of powerful jets and radio lobes. The jet position angle is $150 \pm 10$\degr\ near the core, as obtained with 1.5\arcsec\ VLA maps at 6 and 20 centimeters \citep{Giovannini1994}. The radio lobes are prominent with hot spots and bright outer edges. This means that this is a typical FR-II object.

The lobe-to-core separation ($\sim$ 100~kpc, \citealt{Giovannini1994}) is unfortunately too small to be resolved by optical telescopes at gigaparsec distances, but it remained an interesting exercise to observe it with the HST/FOC. This was done during HST cycle 6 on August 20, 1997, during program ID 6927, using the POL and F342W filters. The F342W filter has a pivot wavelength of $\sim 3404$\AA, and the RMS bandwidth of the filter and detector was $\sim 221$\AA. The observation lasted 1\,940 seconds in total, and the detector mode was 512 $\times$ 1024 pixels in order to zoom into the point source.

As expected, at the native pixel resolution of the instrument (\hbox{0.01435\arcsec$\times$0.01435\arcsec}), we detected no polarization at $\geq 3\sigma_P$ levels (see Fig.~\ref{fig:3C109_FOC} (left)). Thus, the polarization maps were produced using increasing binning and smoothing values until at least one pixel showed a $\geq 3\sigma_P$ polarization measurement. This resulted in a binning of 0.10\arcsec\ and a smoothing of 0.15\arcsec\ FWHM. The total flux map of \object{3C 109} with $\geq 3\sigma_P$ polarization vectors superimposed to it is shown in Fig.~\ref{fig:3C109_FOC} (right).

\begin{figure}[!ht]%
    \centering%
    \includegraphics[height=5cm]{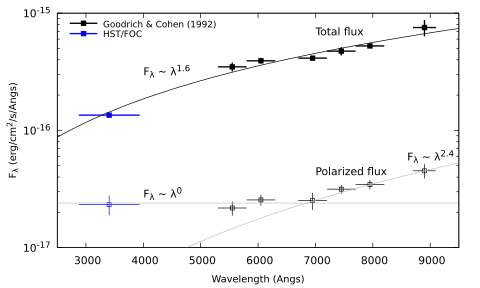}
    \includegraphics[height=5cm]{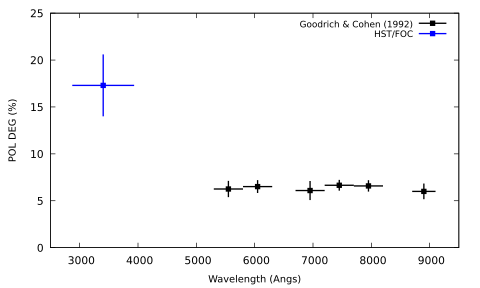}
    \includegraphics[height=5cm]{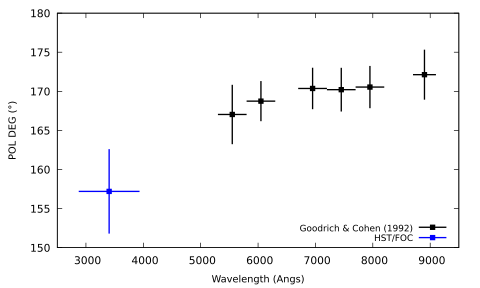}%
    \caption{Broadband polarimetry of \object{3C 109}. The black points are from \citet{Goodrich1992}, and the blue data are from our analysis. All data points were measured using a 2\arcsec\ aperture. The top panel shows the total (filled squares) and polarized (empty squares) intensities as a function of wavelength. Power laws were fit to the total (black) and polarized (gray) intensities. The photon index $\alpha$ of each power law is indicated in the graph. The continuum linear polarization degrees (in percent) are shown in the middle panel, and the associated polarization position angles are shown in the bottom panel.}%
    \label{fig:3C109_spectropolarimetry}%
    \vspace{-6pt}%
\end{figure}

The polarization map we obtained is disappointing, but at least the AGN is not unpolarized. There is only a punctual source with four vectors, and the object no longer shows asymmetry in the total flux. At a higher spatial resolution (Fig.~\ref{fig:3C109_FOC}, left), the polarization information is lost, but the object appears to be slightly elongated along the northeast-southwest direction. The integrated polarization from the full FoV at a low spatial resolution is about \hbox{$P = 20 \pm 2$\%} at \hbox{$\Psi = 158 \pm 3$\degr}. Interestingly, the electric vector polarization angle is perfectly aligned with the radio position angle of the jet, as expected from AGNs seen from the polar direction \citep{Antonucci1993}. We simulated various circular apertures around the flux peak of \object{3C 109} and present the associated photopolarimetric measurements in Tab.~\ref{tab:3C109}. We note that with decreasing aperture radii, the polarization degree decreases while the polarization angle remains constant. At the smallest aperture radius, comparable to the pixel size and Gaussian smoothing, centered on the source and encompassing the whole point-like emission, the constancy of the total flux, polarization degree, and polarization angle with respect to the larger apertures tell us that we do not even start to separate the lobes, jet, and AGN core components.
\begin{table}[!ht]%
    \caption{Observed intensities and polarization from \object{3C 109} at various apertures.}\vspace{-6pt}%
    \centering\resizebox{\linewidth}{!}{%
        \begin{tabular}{l|c c c}
            Aperture & Intensity & $P$ & $\Psi$ \\
            \hline
            FoV (\hbox{7\arcsec$\times$7\arcsec}) & 5.70 $\pm$ 0.07 & 20.1 $\pm$ 2.0 & 158.6 $\pm$ 2.9 \\
            Central \hbox{2\arcsec$\times$2\arcsec} & 1.51 $\pm$ 0.04 & 16.8 $\pm$ 4.3 & 157.1 $\pm$ 7.3 \\
            3\arcsec\ (circular) & 2.08 $\pm$ 0.03 & 20.4 $\pm$ 2.3 & 157.7 $\pm$ 3.2 \\
            2\arcsec\ (circular) & 1.70 $\pm$ 0.02 & 18.5 $\pm$ 2.0 & 157.1 $\pm$ 3.2 \\
            1\arcsec\ (circular) & 1.46 $\pm$ 0.02 & 16.7 $\pm$ 1.6 & 157.6 $\pm$ 3.0 \\
            0.5\arcsec\ (circular) & 1.38 $\pm$ 0.01 & 15.7 $\pm$ 1.4 & 159.0 $\pm$ 2.6 \\
        \end{tabular}%
    }%
    \tablefoot{Intensities are in units of 10$^{-16}$ erg~cm$^{-2}$~s$^{-1}$~\AA$^{-1}$, $P$ are in percent, and $\Psi$ are in degrees.}\label{tab:3C109}%
    \vspace{-6pt}%
\end{table}%
We compared our polarization measurement with those obtained with previous instruments (see Fig.~\ref{fig:3C109_spectropolarimetry}). \citet{Goodrich1992} used a spectropolarimeter attached to the double spectrograph on the Hale 5~m reflector of Palomar Observatory and obtained data between 3420\AA\ and 5140\AA\ and between 5280\AA\ and 10\,120\AA\ in a 2\arcsec\ aperture (slit width). The position angle of the slit on the sky is not known, and the blue data are just vaguely presented in their paper. This prevents us from comparing our 3402\AA\ data point to their near-UV results. At least in the optical and near-IR, they reported a polarization degree of $\sim 6$\% with a polarization position angle of $\sim 170$\degr, that is, not exactly parallel to the radio position angle ($150 \pm 10$\degr). The observed polarization in the broad emission lines allows us to rule out synchrotron radiation as the origin of the continuum polarization \citep{Goodrich1992}. Our near-UV measurement at 2\arcsec\ (see Tab.~\ref{tab:3C109}) has a much higher polarization degree than the optical and near-IR polarization reported by \citet{Goodrich1992}, together with a slightly different polarization angle. This is most certainly due to a combination of 1) a lower contribution of diluting starlight from the host galaxy in the UV with respect to the optical and near-IR bands, and 2) the wavelength dependence of dust scattering and dichroic absorption from the intersecting dust lane. The former is supported by the strong increase of $P$ at UV wavelength, a behavior commonly seen in type-2 AGNs (see, e.g., \citealt{Marin2018}), while the latter is supported by the continuous rotation of the polarization position angle with wavelength, which starts at $\sim 172$\degr\ in the near-infrared and rotates to $\sim 157$\degr\ in the near-UV. The red band is likely dominated by dichroic absorption from the extended dust lane inferred in \object{3C 109}, while the blue band is freer from this contribution and reveals the intrinsic polarization from the AGN. Some polarimetric measurements were also obtained by \citet{Rudy1983,Rudy1984} on the 2.3 m telescope of Steward Observatory in the optical band at an aperture of 2.9\arcsec. They reported that in November 1982, they integrated from 3200\AA\ to 8600\AA, \hbox{$P = 9.31 \pm 2.58$\%} with a polarization angle \hbox{$\Psi = 171 \pm 9$\degr}. This is consistent with our findings: The integrated blue-to-red polarization is higher than what is found in the optical band alone because the higher polarization contributes in the near-UV, which affects the measured polarization angle as well.
To conclude the analysis of \object{3C 109}, we investigate in Fig.~\ref{fig:3C109_spectropolarimetry} (top) the behavior of the total and polarized intensities as a function of wavelength using the data points reported by \citet{Goodrich1992}. The total flux increases with wavelength, following a power law with an alpha index (in lambda) of 1.6. The spectral energy distribution is clearly reddened, which again indicates the existence of a dust lane within the source. The polarized flux, that is, the multiplication of the total flux with the polarization degree, yields a spectrum that is quite different. Not one, but two power laws are required to fit the polarized flux data points. The first power law, from the near-infrared to the optical, has an alpha index (in lambda) of 2.4, that is, much redder than in total flux. In contrast, from the optical to the near-UV, the second power law has an alpha index close to 0, bluer (flatter) than observed in total flux at the same wavelengths. This indicates a transition to a less dust-affected scattering mechanism. The spectral index of $\sim$ 0 implies that a different scattering mechanism takes over, likely electron or Rayleigh-like scattering in the AGN core. In conclusion, the observed transition in polarized spectral shape implies a shift from dust-dominated polarization (at longer wavelengths) to intrinsic AGN polarization (at shorter wavelengths), where electron scattering of the central engine radiation becomes more significant.

\section{Discussion}
\label{sec:Discussion}
Complementary to the polarization analysis of \object{IC 5063} in the first paper of this series \citep{Barnouin2023}, we presented  the two other AGNs (\object{Mrk 3} and \object{Mrk 78}) that belonged to HST program GO 5918. This set of observations was proposed by Axon and collaborators to unveil the origin of the polarization of forbidden lines observed in Seyfert galaxies, and to determine wether the NLR polarization observed in \object{NGC 1068} is typical: an overall low [OIII] emission-line polarization with $P\sim1$\% and few hotspots showing $P$ as high as $P\sim15$\% at a different polarization angle than the UV continuum \citep{Antonucci1985,Inglis1995,Capetti1995a}. As for \object{IC 5063}, observation of the NLR of \object{Mrk 3} and \object{Mrk 78} revealed a low polarization degree ($P\sim1$\%) in the F502M filter, as expected from unpolarized [OIII] doublet emission lines. ISP cannot be the cause of this low polarization, as also demonstrated in the case of \object{IC 5063}, for which we identified interstellar polarization from the host galaxy without significant contribution from the Milky Way \citep[HIP102463, $P<0.1$\% ;][]{Panopoulou2025}. Similarly, in \object{Mrk 78}, we confidently exclude Galactic ISP: The higher polarization degree and the marked difference in polarization angle compared to nearby field stars argue against a foreground origin. Although the polarization remains low ($P\sim1$\%), the angle of \object{Mrk 3} is more consistent with local stellar values, however, which suggests a possible mild contamination from Galactic ISP that cannot be entirely ruled out. Based on this, we confirm that the three observed Seyfert 2 galaxies show similar NLR polarization features as observed in \object{NGC 1068}. Lack of spectropolarimetry in these single-band observations prevented us from estimating the contribution of the [OIII] lines and from computing the underlying continuum polarization.

In the F342W filter, while containing the [NeV] and [OII] emission lines, the UV continuum from the nucleus scattered onto our line of sight dominates and displays strong polarization for \object{Mrk 3} and \object{Cygnus A}of up to $P = 10.1 \pm 0.4$\% and $P = 17.7 \pm 1.5$\% in synthetic apertures centered on their western winds, respectively. These high polarization degrees are associated with polarization angles perpendicular to the respective radio jet axis of each AGN and make a strong case for the diffusion of nuclear light on the NLR. The observed centro-symmetric pattern in the NLR also indicates diffusion from a hidden point source onto our line of sight for \object{Mrk 3}, \object{Mrk 78} and \object{Cygnus A}.

The extension of the observed winds in near-UV allowed us to determine the half-opening angle of the scattering medium and to compare their values to previous imaging and modeling of the NLR for each of these sources. We report a half-opening angle of the biconical winds of $\sim25$\degr\ for \object{Mrk 3} \citep[$\sim22$\degr,][]{Crenshaw2010}, $\sim18$\degr\ for \object{Mrk 78} \citep[$\sim17$\degr,][]{Fischer2011} and $\sim45$\degr\ for \object{Cygnus A} \citep[similar angle observed in][]{Jackson1998}. These values are consistent with the Unified Model for AGN \citep{Antonucci1985}, and similar angles were reported for \object{NGC 1068} and \object{Mrk 463E} \citep{Barnouin2023,Barnouin2024}, for instance. We note that NLRs observed in total intensity usually have larger half-opening angles than NLRs seen in polarized fluxes, but this was not observed for \object{Mrk 3} and \object{Mrk 78} through filter F502M because the dominant contribution comes from the unpolarized [OIII] emission lines. \object{Cygnus A} observation lacks the statistics for us to conclude on the extension of the ionized winds in polarized flux.

While proposal ID 3790 searched for reflected nuclear light surrounding radio galaxies including \object{Cygnus A}, proposal IDs 6510 and 6702 aimed to spatially resolve and characterize off-nuclear scattering mirrors in two colors (F275W and F342W) for \object{Cygnus A} and \object{Mrk 3}. The results presented here show the unmatched capabilities of near-UV high-resolution imaging polarization to successfully conduct a study like this.

Finally, the high-resolution polarization maps of \object{Cygnus A} showed a clear centro-symmetric pattern in the UV and optical \citep{Jackson1998}. Based on this geometrical constraint, we should be able to locate its obscured nucleus in order to properly correlate the radio, infrared, optical, and UV observations. This is of prime importance for multiwavelength campaigns that try to correlate supermassive black hole activities with large-scale physical processes \citep{Ogle1997,Hurt1999,Lopez-Rodriguez2018,Lo2021}. Despite the great quality of the \object{Cygnus A} observation we presented, we lack the statistics to perform this study. Further imaging polarimetry observations are required.

\section{Conclusion}
\label{sec:Conclusion}
Our conclusions for the objects are listed below.
\begin{itemize}
    \item The polarized observations of \object{Mrk 3} revealed an S-shaped structure that is aligned with the radio jets and a strong east-west polarization asymmetry in the NLR. This suggests anisotropic scattering affected by geometry and local extinction, with a reversal of this asymmetry from the optical to the UV.
    \item We detected for the first time a UV polarization from \object{NGC 3862} at more than 3$\sigma$. The moderate but significant polarized signature indicates a compact synchrotron origin that is aligned with the base of the radio jet and diluted by the host starlight.
    \item Our polarized observations of \object{Mrk 78} revealed a highly polarized knot ($P$ = 11.8\%) north of the southwest hotspot. This is consistent with a scenario of perpendicular nuclear light that scatters on the NLR and confirms the biconical geometry of the emissive gas. It also highlights jet-gas interactions at different evolutionary stages.
    \item The first near-UV polarized maps of \object{Cygnus A} revealed a centrosymmetric polarization pattern (with P reaching 51\% in a single spatial bin) in the southeast and northwest regions, consistent with light scattering from a hidden nucleus, with a dominant component originating from the northwest hotspot.
    \item  High polarization degrees (about 20\%), associated with a polarization position angle parallel to the radio jet axis, were found in the broad-line radio galaxy \object{3C 109}. The analysis of the polarization of \object{3C 109}, in particular, its different total and polarized intensities, provides strong evidence for a dust-obscured quasar-like nucleus in which multiple scattering components contribute to its observed polarization properties.
\end{itemize}
Our study thus revealed significant polarization detections in all five of our targets. For three of them, the polarization angle is perpendicular to the radio axis, and the observed high polarization is consistent with a perpendicular scattering mechanism from an obscured nuclear source. In the other two cases, the alignment of the polarization with the jet radio axis indicates either a dominant nonthermal component (\object{NGC 3862}) or scattering from an equatorial medium viewed from above (\object{3C 109}).

These results demonstrate the potential of HST/FOC archives when processed with modern recalibration and reduction methods. They pave the way for the widespread use of high spatial resolution UV polarimetry as a diagnostic tool for AGNs.

\section*{Data availability}
All reduced datasets presented in this paper are made available in standard FITS file form at the CDS via anonymous ftp to cdsarc.u-strasbg.fr or via \url{https://cdsweb.u-strasbg.fr/cgi-bin/qcat?/A+A/}.

\begin{acknowledgements}
    The authors thanks the referee for their comments that helped improve this paper. TB and FM acknowledge the support of the CNRS, the University of Strasbourg, the ATPEM and ATCG. This work was supported by the "Action Th\'ematique des Ph\'enom\`enes Extr\^emes et Multi-messagers" (ATPEM) and the "Action Th\'ematique de Cosmologie et Galaxies (ATCG)" of CNRS/INSU co-funded by CNRS/IN2P3, CNRS/INP, CEA and CNES. E.L.-R. thanks support by the NASA Astrophysics Decadal Survey Precursor Science (ADSPS) Program (NNH22ZDA001N-ADSPS) with ID 22-ADSPS22-0009 and agreement number 80NSSC23K1585.
\end{acknowledgements}
\bibliographystyle{aa}
\bibliography{biblio}
\begin{appendix}
\onecolumn
\section{\object{Mrk 3}}%
\begin{figure*}[!ht]
    \centering%
    \includegraphics[height=6.5cm]{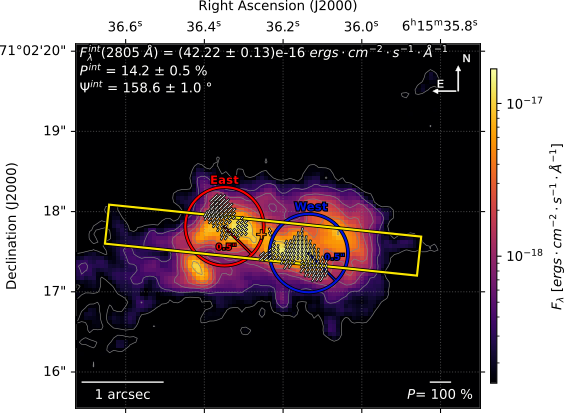}\hspace{3mm}%
    \includegraphics[height=6.5cm]{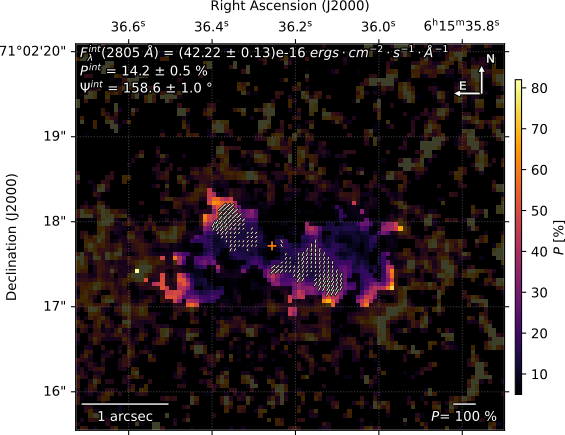}%
    \caption{HST/FOC polarization maps of \object{Mrk 3} (observation ID 6702) through filter F275W. \textit{Left:} Intensity map overlaid with integration regions used in Table~\ref{tab:Mrk3}. \textit{Right:} Debiased polarization degree. Darker pixels have $\left[S/N\right]_P < 1$. Both maps are shown at a spatial binning of 0.05\arcsec, smoothed using a 0.075\arcsec FWHM Gaussian kernel. 176 pixels have $\left[S/N\right]_P \geq 3$.}\label{fig:MRK3_6702_F275W}%
\end{figure*}

\begin{figure*}[!ht]
    \centering%
    \includegraphics[height=6.5cm]{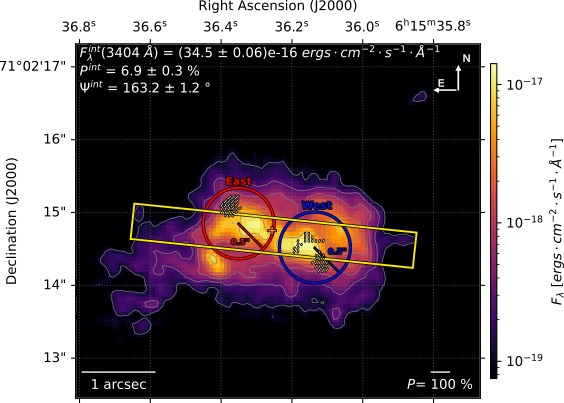}\hspace{3mm}%
    \includegraphics[height=6.5cm]{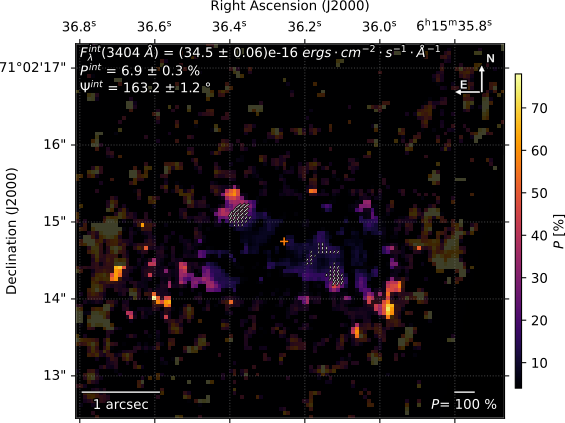}%
    \caption{HST/FOC polarization maps of \object{Mrk 3} (observation ID 6702) through filter F342W. \textit{Left:} Intensity map overlaid with integration regions used in Table~\ref{tab:Mrk3}. \textit{Right:} Debiased polarization degree. Darker pixels have $\left[S/N\right]_P < 1$. Both maps are shown at a spatial binning 0.05\arcsec, smoothed using a 0.075\arcsec FWHM Gaussian kernel. 63 pixels have $\left[S/N\right]_P \geq 3$. The orange cross displays the estimated location for the obscured nucleus \citep{Kishimoto2002}.}\label{fig:MRK3_6702_F342W}%
\end{figure*}

\label{app:Mrk3}
To complete the near-UV polarization study of \object{Mrk 3} in Sect.~\ref{subsec:Mrk3}, we reduced observation ID 5918 through filter F502M alongside previously published observations through filters F275W and F342W \citep{Kishimoto2002}. While the former study the [OIII] polarization, the latter focuses on the continuum polarization resulting from the diffusion of nuclear light onto our line of sight by the NLR. We present in Fig.~\ref{fig:MRK3_6702_F275W} and Fig.~\ref{fig:MRK3_6702_F342W} the polarization maps obtained with our reduction pipeline for observation through filter F275W and F342W, respectively. Images have been resampled to 0.05\arcsec$\times$0.05\arcsec\ pixels and smoothed with a Gaussian with an FWHM 0.075\arcsec, that is, $1.5$ times the size of a resampled pixel. Those polarization maps benefit for the latest MAST calibration and are presented here at higher spatial resolution than previously published in \cite{Kishimoto2002}. The polarization components integrated on synthetic apertures displayed on top of the total intensity maps (left) are compared to the [OIII] polarization in Table~\ref{tab:Mrk3} and discussed in Sect.~\ref{subsec:Mrk3}.


\section{Cygnus A}%
\begin{figure*}[!ht]%
    \centering%
    \includegraphics[height=7cm]{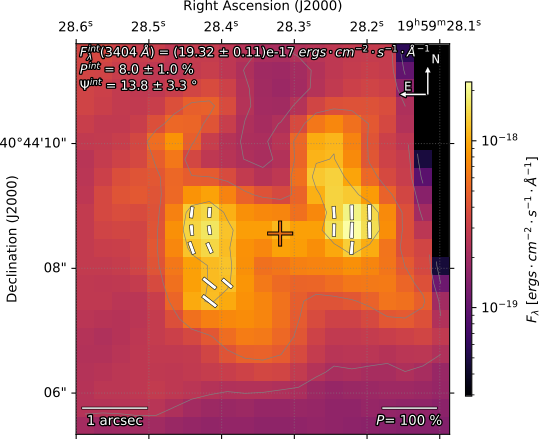}\hspace{3mm}%
    \includegraphics[height=7cm]{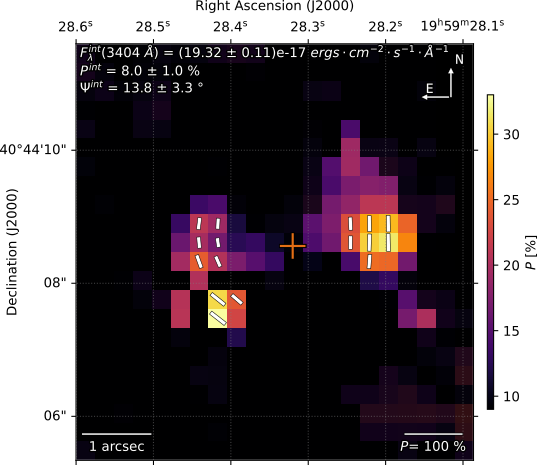}%
    \caption{HST/FOC polarization maps of \object{Cygnus A}, observation ID 6510 through filter F342W. \textit{Left:} Intensity map. \textit{Right:} Debiased polarization degree. Both maps were resampled to 20px$\times$20px (corresponding to 0.28\arcsec$\times$0.28\arcsec), smoothed with a Gaussian with an FWHM $1.5$ times the size of the resampled bin size (0.42\arcsec). 16 pixels have $\left[S/N\right]_P \geq 3$. The orange cross displays the estimated location for the obscured nucleus \citep{Jackson1998}.}\label{fig:CygnusA6510_largebins}%
\end{figure*}

\label{app:CygnusA}
To complete the high resolution near-UV imaging polarization study of \object{Cygnus A}, we resampled the obtained polarization maps to large spatial bins in order to study to obtain enough statistics further from the nucleus. We present in Fig.~\ref{fig:CygnusA6510_largebins} the polarization maps with spatial bins of 0.28\arcsec$\times$0.28\arcsec\ (resampling to 20$\times$20 native pixels). The obtained polarizations maps shows 16 spatial bins for which we detect polarization ($\left[S/N\right]_P \geq 3$), arranged in a centro-symmetric pattern. Normals to the detected polarization electric vectors concurs near the estimated location of the hidden nucleus \citep{Jackson1998}, consistent with light scattering from this obscured point source. The polarization degree range from 15\% to 20\% in both winds, reaching 30\% in the polarized hotspots observed at higher resolution. The polarization degree map is put as a reference for future studies of the winds of \object{Cygnus A}, the detailed polarization maps are available on demand and will be rendered public among the full catalog of AGNs in the polarized archives of the HST/FOC.

\begin{figure*}[!ht]
    \centering%
    \includegraphics[height=9cm]{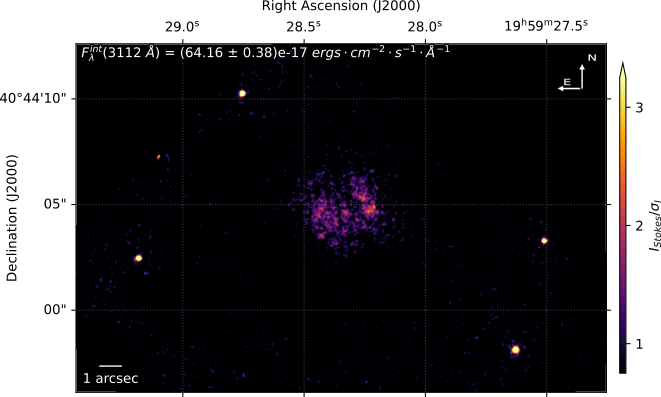}%
    \caption{HST/FOC $\left[S/N\right]_I$ map of observation ID 3790 through filter F320W of \object{Cygnus A}. FoV shows five point-like sources that seem arranged in a point-symmetry fashion around the AGN.}\label{fig:CygnusA3790_lowflux_srcs}%
\end{figure*}

We report, in the observation of \object{Cygnus A} through filter F320W ($\lambda\sim3112$\AA), five point-like sources that seem symmetrically arranged around the central AGN. For better contrast in this pre-COSTAR and low exposure observation, we show in Fig.~\ref{fig:CygnusA3790_lowflux_srcs} the S/N per spatial pixel in total intensity. Due to the apparent symmetry around a known massive object such as \object{Cygnus A}, we timidly put forward the hypothesis that these sources might result from strong lensing of a background source by the AGN. They are observed in near-UV and we found counterparts in the near-infrared \citep[R-band imaging:][Fig.~1]{Nilsson1997}. Polarization may arise from gravitational lensing, but the HST/FOC observation reduced and analyzed here lacks the statistics for such study. Further spectroscopic and polarimetric observations are required to better understand the nature of these objects (see Marin et al., in prep.).

\begin{figure*}[!ht]
    \centering%
    \includegraphics[height=10cm]{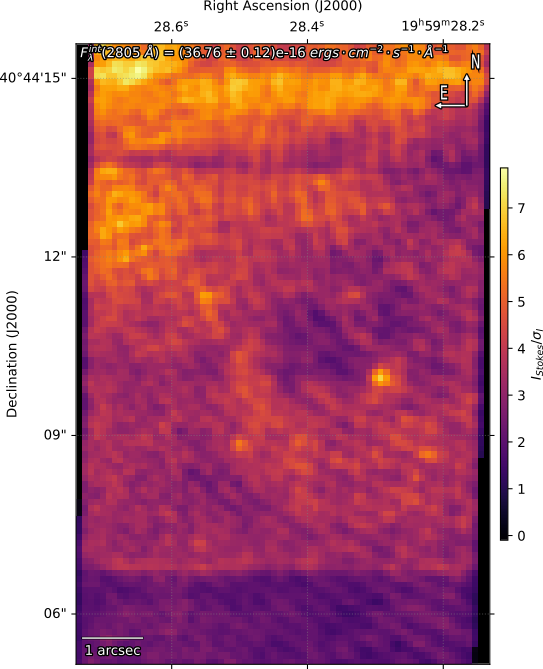}\hspace{3mm}%
    \includegraphics[height=10cm]{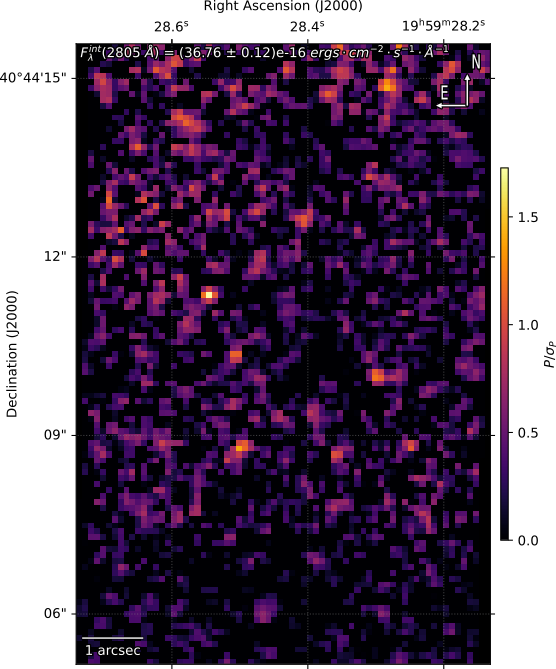}%
    \caption{Signal-to-noise maps of the underexposed HST/FOC observation of Cygnus A, observation ID 6510 through filter F275W. \textit{Left:} $\left[S/N\right]_I$. \textit{Right:} $\left[S/N\right]_P$.}\label{fig:CygnusA6510_lowflux}%
\end{figure*}

In the sake of completeness, we also publish the maps of both S/N in total intensity and polarization degree for observation ID 6510 through filter F275W of \object{Cygnus A} in Fig.~\ref{fig:CygnusA6510_lowflux}. This observation is well under-exposed and no analysis of this dataset could be performed.

\end{appendix}
\end{document}